\documentstyle[prd,aps,preprint,tighten]{revtex}
\def\thefootnote{\fnsymbol{footnote}}
\def\Tr{{\rm Tr}}
\def\re{{\rm Re}}
\def\im{{\rm Im}}
\def\[{\left [}
\def\]{\right ]}
\def\({\left (}
\def\){\right )}
\def\lbr{\left\{}
\def\rbr{\right\}}

\def\tG{{\tilde G}}

\def\gh{{\hat G}}

\def\T{\bar{T}}
\def\t{\bar{t}}

\def\bp{\bar{\phi}}

\newcommand{\lowest}{_{\theta =\bar{\theta}=0}}

\newcommand\be{\begin{equation}}
\newcommand\ee{\end{equation}}
\newcommand\bea{\begin{eqnarray}}
\newcommand\eea{\end{eqnarray}}

\newcommand\pa{\partial}

\newcommand\eq[1]{Eq.~(\ref{#1})}

\newcommand{\sub}[1]{_{\mbox{\scriptsize#1}}}
\newcommand{\su}[1]{^{\mbox{\scriptsize#1}}}

\newcommand\GeV{\,\mbox{GeV}}
\newcommand\TeV{\,\mbox{TeV}}
\newcommand\mpl{M_{\rm Pl}}

\def\lsim{\lesssim}

\tighten

\begin{document}
\begin{titlepage}
\pagestyle{empty}
\begin{center}
\tighten
    \hfill LANCS-TH/9705 \\
      \hfill  LBNL-40830 \\ 
      \hfill  UCB-PTH-97/48 \\
      \hfill hep-th/9806157\\
    \hfill \today 
\vskip .2in
{\large \bf Inflation and flat directions in modular invariant superstring 
effective theories}\footnote{
This work was supported in part by 
PPARC grant GR/L40649 and NATO 
grant CRG 970214, 
in part by 
the Director, Office of Energy Research, Office of High Energy and Nuclear
Physics, Division of High Energy Physics of the U.S. Department of Energy under
Contract DE-AC03-76SF00098 and in part by the National Science Foundation under
grant PHY-95-14797.  HM was also supported by Alfred P. Sloan 
foundation.} 
\vskip .2in
\vskip .2in
Mary K. Gaillard$^{\S\dag}$, David H. Lyth{}$^{\ddag}$ {\em and}\/
{Hitoshi Murayama$^{\S\dag}$}
\vskip .2in

{}$^{\ddag}${\em Department of Physics, Lancaster University, 
Lancaster, LA1 4YB, U.K.}

{}$^{\S}${\em Department of Physics, University of California, 
 Berkeley, California 94720 }

{}$^{\dag}${\em Theoretical Physics Group, Earnest Orlando Lawrence
 Berkeley National Laboratory, Berkeley, California 94720 }

\end{center}

\tighten

\begin{abstract}

 The potential during inflation must be very flat in, at least, the direction 
of the inflaton. In renormalizable global supersymmetry, flat directions 
are 
ubiquitous, but they are not preserved in a generic supergravity theory. It 
is known that at least some of them are preserved in no-scale supergravity, 
and simple generalizations of it. We here study a more realistic 
generalization, based on string-derived supergravity, using the linear 
supermultiplet formalism for the dilaton. We consider
a general class of hybrid inflation models, where 
a Fayet-Illiopoulos $D$ term drives some fields to large values.
The potential is dominated by the $F$ term, but 
flatness is preserved in some directions. This allows inflation, with
the dilaton stabilized in its domain of attraction, and 
some moduli stabilized at their vacuum values. 
Another modulus may be the inflaton.

\end{abstract}
\end{titlepage}

\newpage
\renewcommand{\thepage}{\roman{page}}
\setcounter{page}{2}
\mbox{ }

\vskip 1in

\begin{center}
{\bf Disclaimer}
\end{center}

\vskip .2in

\begin{scriptsize}
\begin{quotation}
This document was prepared as an account of work sponsored by the United
States Government. While this document is believed to contain correct 
 information, neither the United States Government nor any agency
thereof, nor The Regents of the University of California, nor any of their
employees, makes any warranty, express or implied, or assumes any legal
liability or responsibility for the accuracy, completeness, or usefulness
of any information, apparatus, product, or process disclosed, or represents
that its use would not infringe privately owned rights.  Reference herein
to any specific commercial products process, or service by its trade name,
trademark, manufacturer, or otherwise, does not necessarily constitute or
imply its endorsement, recommendation, or favoring by the United States
Government or any agency thereof, or The Regents of the University of
California.  The views and opinions of authors expressed herein do not
necessarily state or reflect those of the United States Government or any
agency thereof, or The Regents of the University of California.
\end{quotation}
\end{scriptsize}

\vskip 2in

\begin{center}
\begin{small}
{\it Lawrence Berkeley Laboratory is an equal opportunity employer.}
\end{small}
\end{center}

\newpage
\renewcommand{\thepage}{\arabic{page}}
\setcounter{page}{1}
\def\thefootnote{\arabic{footnote}}
\setcounter{footnote}{0}
\section{Introduction}

Cosmological inflation has been regarded as the most elegant solution 
to the horizon and flatness problems of the standard Big Bang 
universe.  Even though it explains  why the current 
Universe appears so homogeneous and flat in a natural manner, it 
has been difficult to construct a model of inflation without a small 
parameter.  In fact, one needs a scalar field (inflaton) that rolls 
down the potential very slowly to successfully generate a viable
inflationary scenario \cite{LL2,nrev}.
This requires the potential to be 
almost flat in the direction of the inflaton.

Other cosmological considerations may call for a flatness of the 
potential in non-inflaton directions. 
For instance, the Affleck--Dine 
baryogenesis scenario \cite{affdine1} requires a scalar field which 
carries baryon number to have a large amplitude to start with.  To 
maintain a large amplitude during the rapid expansion of the universe, 
the scalar potential needs to either be almost flat
\cite{affdine1,cgmo}, or to have a 
negative squared mass \cite{affdine2,ewanbg}.\footnote
{It should be emphasized, though, that we are looking at 
the global SUSY flat
directions only {\em during} inflation.  After inflation is over
one has rapidly oscillating fields and/or thermalized particles, 
and a separate discussion is required in this much more complicated
situation.  One might argue 
\cite{affdine2}
that all fields are likely to 
acquire masses at least of order $H$ (until $H$ falls below their true 
mass).  If that is so, Affleck-Dine
baryogenesis may proceed more or less as in \cite{affdine2}
or \cite{ewanbg}, whether or not the directions responsible for it are 
flat during inflation.  However, the thermal effects are
exponentially suppressed if the amplitude of the scalar field is
larger than the temperature.  The effect of the oscillating field on
the flat direction depends sensitively on the structure of the
non-renormalizable K\"ahler potential terms.  Discussion on these
issues is beyond the scope of this paper.}

In general, it 
appears rather unnatural to impose an almost flat scalar potential
in quantum field theory, because such flatness is likely to be
destroyed by radiative corrections.  Specifically, scalar field masses are
generally quadratically divergent and are not protected
by any symmetries.\footnote{The exception is when the scalar field is
a Nambu--Goldstone boson of a spontaneously broken global symmetry.}
Supersymmetry, however, may maintain the flatness of a 
 tree-level scalar 
potential against radiative corrections due to the nonrenormalization
theorem.

A renormalizable, globally supersymmetric theory typically 
has several directions in which the tree-level potential is very flat.
However, because inflation couples the 
energy density of a scalar potential to gravity to cause a rapid 
cosmological expansion, supersymmetry has to be made local: 
supergravity.  During inflation,
a generic supergravity theory 
lifts the flat directions of global supersymmetry, generating 
\cite{coughlan,cllsw} a
squared mass at least of order $3H^2\simeq \mpl^{-2}V$ in magnitude.\footnote
{Here, $H$ is the Hubble parameter defined in terms of the scale factor 
$a$ by $H=\dot{a}/a$, and 
$\mpl\equiv(8\pi G)^{-1/2}
=2.4\times 10^{18}\GeV$ is the reduced Planck mass. The inflaton 
potential is $V(\phi)$, a prime denotes
$d/d\phi$, and we have remembered that $V\simeq 3\mpl^2H^2$ during 
inflation.}

This generic result 
must be evaded for the
inflaton field, \cite{cllsw}
since in its direction $|V''|\ll \mpl^{-2}V$ is necessary
for slow-roll inflation. ($V''$ is the second derivative of the inflaton 
potential, which cannot be much less than the squared mass 
along the inflaton trajectory.)
One would like to understand how this evasion comes about, and
whether it occurs for scalar fields other than the inflaton.

The question of whether a scalar potential has a flat direction consistent 
with inflation requires knowledge not only of the superpotential, 
which is not renormalized so that a specific form of it is at least 
technically natural, but also the K\"ahler potential, which can be
renormalized by higher dimension operators with arbitrary coefficients in
generic supergravity.  In fact, one needs to know the K\"ahler 
potential at least up to quartic terms to determine if the potential is flat. 
Since these contributions are arbitrary in general supergravity, a natural 
question is whether an underlying quantum theory of gravity, such as 
superstring theory, determines a specific form of K\"ahler potential that 
ensures the flatness of the scalar potential along certain directions in the 
field space.   

This question has been a difficult one to
address, because of other cosmological problems in superstring-inspired
supergravity theories.  The dilaton field exhibits a runaway behavior
and it has been difficult to obtain a minimum of the
potential, consistent with spontaneously broken supersymmetry (as
required by phenomenology) and vanishing cosmological constant (at
least on the scale of supersymmetry breaking).  Recently, a modular invariant
formalism, based on string orbifold compactification, was proposed to study the 
stabilization of the dilaton by employing the linear
multiplet description of the dilaton \cite{bgw}.  

In the present paper, we explore the possibilities for inflation
in the context of this formalism.
Because we are writing down the supergravity Lagrangian after integrating
out all the massive string and Kaluza--Klein excitations around a
consistent vacuum, the superpotential has a power series expansion in 
the matter fields. It 
starts at the cubic order,
and
higher order terms are allowed with power suppression in the string 
scale. The effective mass terms and/or linear terms that are presumably 
necessary for inflation will appear when some of the fields acquire 
nonzero values ($vev$'s).\footnote
{Other intermediate-scale $vev$'s occurring in Nature (associated say with 
Peccei-Quinn symmetry, neutrino masses or a GUT) might be related to
the inflationary one, or they might be in a different sector of the theory.
}
We suppose, following Stewart \cite{ewansg},
that the $vev$'s are generated when 
a Fayet-Illiopoulos $D$ term is driven to a small value.
We show how this can 
preserve some of the flat directions of global SUSY,
generating a potential flat enough for inflation, which will probably be 
of the hybrid \cite{LIN2SC,cllsw} type. 
(During hybrid inflation a non-inflaton field is 
displaced from the vacuum, 
and is responsible for most of the potential.)

Our paper is organized as follows.  In the next section, we introduce 
the special form of the supergravity Lagrangian obtained from superstrings
in \cite{bgw}.  Section 3 contains the strategy for inflation 
model-building, and the main discussion on the flatness of the scalar 
potential. We conclude in Section 4. 

We generally set $\mpl=1$, where $\mpl=(8\pi G)^{-1/2}
=2.3\times 10^{18}\GeV$.

\section{Superstring-derived supergravity}

We are going to show how to construct models of inflation, in the
context of a realistic supergravity theory derived from the weakly
coupled superstring \cite{bgw}. It is based on a class of 
orbifold compactifications\cite{iban,font2} with three untwisted three moduli 
$t_I$, and contains an effective potential for the dilaton, induced by gaugino 
condensation. Dilaton stabilization in the true vacuum is achieved by the 
inclusion of nonperturbative string effects in the K\"ahler potential, that 
modify the form of this potential.\footnote{In this paper ``modulus''
refers to the three untwisted moduli of the class of orbifold compactifications
that we consider in explicit examples.  In the usual chiral formalism, the
dilaton and the universal axion are the real and imaginary parts of of a complex
field $s$. In the linear multiplet formalism used here, the axion is replaced
by a two-form potential $b_{\mu\nu}$ related to $\im s$ by a duality 
transformation that determines the dilaton $\ell$ in the classical limit
as $\ell = (2{\rm Re} s)^{-1}$; this relation is modified in the presence of 
both perturbative and nonperturbative quantum effects.}

An important part of our program is to demonstrate that the dilaton can
be stabilized during inflation, by the same nonperturbative string effects
that stabilize it in the true vacuum. First though, we look at
a simplified model which ignores the dilaton and the gaugino condensate. 
Then the scalar fields are all in chiral multiplets; they consist of the
moduli $t_I$, and 
matter fields which we shall denote by $\phi_\alpha$.\footnote
{We include as ``matter'' the so-called twisted
moduli that are Standard Model gauge singlets, but have nonvanishing modular 
weights. For our purposes, their couplings are no different from those of 
twisted matter fields that are SM gauge nonsinglets.} 

The tree-level potential has the usual form $V=V_F + V_D$. The $F$ term is
\be
V_F= e^K\left[\sum_{nm} K^{n\bar m}\left(W_n + K_n W\right)
\left(\bar W_{\bar m}+K_{\bar m} \bar W\right)
-3|W|^2 \right] \,.
\label{1} \ee
The superpotential $W$ is a holomorphic function of the 
complex scalar fields, while the K\"ahler potential
$K$ is a function of the fields and their
complex conjugates. A subscript $n$ denotes the derivative
with respect 
to the $n$th field, and $ \bar n$ the derivative with respect to 
its complex conjugate. (In this context, $n$ runs over both the
both matter fields and the moduli.) The matrix $K^{n\bar m}$ is the 
inverse of the matrix $K_{n\bar m}$.

\subsection{The potential ignoring the dilaton}

We suppose that the only relevant part of the $D$ term
involves a $U(1)$ with a Fayet-Illiopoulos term,
\be V_D = {g^2\over2}\left(\sum_n q_n K_n \phi_n + \xi_D \right)^2 
\label{dterm98}\,.\ee
In this expression, $n$ runs only over matter fields
charged  under the relevant $U(1)$. Its gauge coupling is $g$, and 
$q_n$ are the charges.
As discussed later, weakly coupled string theory predicts
that $\xi_D$ will be an order of magnitude or so below the Planck scale. 

To warm up, we consider only a single modulus $t$, corresponding to 
compactification 
on a six-torus \cite{witten85b}.
Its K\"ahler potential is
$K=-3\ln x$ where
$x\equiv t+\bar t- \sum_n|\phi_n|^2$, and 
$W$ is independent of $t$. 
This leads to what is termed a 
`no-scale' theory \cite{noscale}, in which \eq{1} becomes simply
(for $|\phi_n|\ll 1$)
\be V=\frac3{x^2}\sum|W_n|^2 \,. 
\label{onet}
\ee

Instead of Re\,$t$, one can regard $x$ as a field since 
this choice too corresponds to approximately canonical normalization.
The precise form of the kinetic terms is given for example in \cite{gmo}.

It looks as if $x$ will run away to infinity, but we shall now see that 
this need not happen if vevs for the matter fields are generated from 
a Fayet-Illiopoulos $D$ term.\footnote
{The possibility of generating vevs from such a term has been considered 
in two previous works 
\cite{cllsw,ewansg}, but these failed to notice 
the crucial effect of the
nontrivial kinetic term $K_{n}$ in \eq{dterm98}.}

Suppose that there is only one $W_n$, say $W_3$, which comes from a 
term $\lambda\phi_1\phi_2\phi_3$ in the power series expansion of $W$,
and occurs because the 
$D$ term lifts the flat 1 and 2 directions. With matter fields
$|\phi_n|\ll 1$ one has $K_{n\bar m}\simeq x^{-1}\delta_{nm}$
so this will generate a vev
\be |\phi_1\phi_2|^2 = c x^2 \xi_D^2 \,,\ee 
where $c$ is a constant of order 1. 
This will give
\be V\simeq \frac3{x^2} |W_3|^2 \simeq 3 c\lambda^2 \xi_D^2 \,.\ee
All of the flat directions are preserved except for $n=1,2$ and $3$, 
and the potential is also flat 
in the direction $t$. Slopes in these
directions can be generated from nonrenormalizable terms, 
departures from the 
no-scale assumption, gaugino condensation or loop effects.
When they are included the inflaton (corresponding to the direction of 
steepest descent in the space of the flat directions)
might turn out to be any combination of
the flat directions and $t$.

Earlier authors working with the potential \eq{onet} supposed instead
that $x$ was fixed, either by an 
ad hoc functional form for $K(x)$ \cite{andreibook,keithrep,hitinf}, 
or by a loop correction
\cite{gmo}. Then all of the flat directions are preserved, and 
Im\,$t$ and $x$ are also flat.

For the rest of this paper, we invoke 
three moduli fields $t_I$.
Also, we allow $W$ to have a dependence on the moduli,
that is determined by the modular invariance of the theory.
The matter fields are divided into
twisted fields $\phi_A$ and untwisted fields $\phi_{AI}$.
For the most part we suppose that the twisted fields
vanish during inflation. 
Ignoring both them and the dilaton, 
the K\"ahler potential is 
\be K= -\sum_{I=1}^{3} \ln x_I , \ee
where 
\be x_I= t_I + \bar t_I - \sum_A|\phi_{AI}|^2 . \ee
The potential \eq{1} becomes
\be V=e^K\left[\sum_I \left(x_I \sum_A
|W_{AI} + \bar \phi_{AI} W_I|^2 +
| x_I W_I - W|^2 \right) - 3|W|^2 \right] . \label{9} \ee
In this expression, $W_I\equiv \pa W/\pa t_I$. 

If $W$ does not depend on the moduli $t_I$, we have simply
\be V=e^K\sum_{I,A} x_I |W_{AI}|^2 \label{1098}\ee
This is the same as in global SUSY, except for the factors
$e^K x_I= x_I/x_1 x_2 x_3$. 

As in the previous case,
a Fayet-Illiopolos $D$ term may prevent the runaway of the
$x_I$, even if $W$ has no dependence on $x_I$.

As pointed out in~\cite{gmo}, the preservation of flat directions for 
the toy models (\ref{onet}) and (\ref{1098}) is a consequence of the 
Heisenberg invariance~\cite{heis} of the K\"ahler potential which 
depends on the scalar fields of these models only through the invariant
combinations $x$ and $x_I$ of the moduli $t$ or $t_I$ and the 
(untwisted) matter fields $\phi_A$ or $\phi_{IA}$,

\subsection{The full potential from orbifold compactification}

To construct a realistic string-derived model, we have to include the dilaton,
the Green-Schwarz term needed to cancel the modular anomaly induced by field
theory loop corrections, a superpotential for the twisted sector fields, and
the effective potential for the dilaton that is
induced by gaugino condensation. The last two terms break Heisenberg invariance
through an explicit dependence on the moduli (although the last will be
considered negligible in much of our discussion). The K\"ahler potential and 
the G-S term are not completely known. In addition to imposing modular 
invariance we will assume that they are Heisenberg invariant.  This is 
equivalent
to the, at least plausible, assumption that the K\"ahler potential and the G-S 
term involve the untwisted scalar fields only through the radii $R_I$
of compactification of the three tori: in string units $1/2R^2_I = 
t_I + \t_I - \sum_A|\phi_{AI}|^2$.   We will indicate the modifications that
occur if this assumption is relaxed.

To incorporate the dilaton, we turn to the model of \cite{bgw}, which arguably
reflects string constraints more faithfully than any so far.
It was realized some time ago~\cite{bdqq,bgt,sduality} 
that the usual chiral formalism for gaugino
condensation, that uses an unconstrained chiral multiplet as in interpolating
field for the gaugino condensate, is inconsistent with the Bianchi identity
for the Yang-Mills chiral superfield. The most straightforward way to introduce
a chiral field with the correct constraint is to identify it with
the chiral projection of a
vector superfield whose components also contain a those of a linear
supermultiplet, interpreted as the dilaton supermultiplet in this formalism.
In fact, the chiral multiplet for the dilaton that is commonly used is obtained
after a duality transformation from the components of a linear supermultiplet
that are remnants of the dilaton, dilatino, and a three-form field of ten
dimensional supergravity.  There is increasing evidence~\cite{lin} 
that the linear supermultiplet is the correct formulation for the dilaton in the
context of superstring theory.  Although it has been argued~\cite{bdqq} that 
the linear and chiral multiplets are equivalent through a duality transformation
even at the quantum level and in the presence of nonperturbative effects, the
implementation of the correct constraint leads to considerable complication in
the chiral formalism.  Using the linear formalism for gaugino
condensation~\cite{bdqq,bgt}, together with constraints
from string theory, including perturbative modular invariance~\cite{mod} and
matching conditions~\cite{gt,kl} at the string scale, as well as a 
parameterization~\cite{bgw1,casas} of nonperturbative string
effects~\cite{shenk} to stabilize the dilaton, it 
was shown~\cite{bgw} that the moduli are naturally stabilized 
at their self dual
points in realistic theories~\cite{iban,font2} from orbifold compactification.
Since the dilaton is stabilized at weak coupling, one expects the result for 
moduli $vev$'s to hold up to possible small corrections that could arise if 
the nonperturbative string
corrections to the dilaton K\"ahler potential are moduli-dependent~\cite{eva}.

We take the K\"ahler potential $K$ and the Green-Schwarz term $V_{GS}$ to be 
\bea K &=& G + \ln V + g(V), \quad G = \tG + \sum_AX_A, 
\quad V_{GS} = b\tG + \sum p_AX_A, \nonumber \\ \tG &=& \sum_I\tG_I, \quad
\tG_I = - \ln(T_I + \T_I - \sum_A|\Phi_{AI}|^2), \quad X_A =
 \exp{\left(\sum_I q^A_I\tG^I\right)}|\Phi_A|^2 , \label{functs}\eea
where $g(V)$ parameterizes nonperturbative string effects,
$b=30/8\pi^2,\;V$ is a  vector superfield whose scalar component
$V_{\lowest} = \ell$ is the dilaton. The $T_I$ are the chiral
multiplets containing the moduli.  The $\Phi_{AI}$ are
untwisted sector chiral multiplets,  $\Phi_{AJ}$ having 
modular weight $q^{AJ}_I= \delta^J_I$, and the $\Phi_A$ are 
twisted sector chiral multiplets
with modular weights $q^A_I>0$ (typically less than 1).\footnote
{The group of modular transformations on the moduli is generated by
$t_I\to 1/t_I$ and $t_I\to t_I\pm i$. Under this group, a matter 
field $\phi_\alpha$ transforms like $\prod_I\eta^{-2q^\alpha_I}(t_I)$,
where $\eta$ is the Dedekind function and $q^\alpha_I$ are the
weights of the field.}
The coefficients $p_A$ are unknown, but
can plausibly be either zero or equal to $b\approx .38$.   If the twisted
sector fields decouple from the GS term we have $p_A=0$, while if the GS term is
simply $V_{GS} = bK$, we have $p_A=b$.  These unknown couplings determine the 
soft
SUSY breaking parameters at the vacuum and also during inflation, and therefore
may be relevant to the issue of maintaining flat directions during inflation.  

To obtain the $K$ that is to be used
in calculating the potential, one replaces $V$, $T_I$, $\Phi_A$ and
$\Phi_{AI}$ by the corresponding scalar fields 
$\ell$, $t_I$, $\phi_A$ and $\phi_{AI}$. (The matter fields will be
denoted collectively by $\phi_\alpha$, with $\alpha=A$ or $AI$.) 
As before, we define $x_I\equiv t_I+\bar t_I-\sum_A|\phi_{AI}^2|$, and we 
also define
\be X_A \equiv  \(\prod_I x_I^{-q^A_I} \)|\phi_A|^2 \,.\label{xa}\ee
Then
\be K= \ln(\ell) + g(\ell) -\sum_I \ln x_I + \sum_A X_A \,,\label{k98}\ee

The choices for $K$ and $V_{GS}$ are consistent with what is known~\cite{acc} 
from string theory.  The Lagrangian for the untwisted sector of the theory 
can be obtained by direct compactification of ten-dimensional supergravity.  
Therefore, we know the part of $K$ that depends only on the
{\em untwisted} fields, which will allow us to keep their masses under
control and hence preserve some untwisted flat directions.  To obtain
information about twisted sector couplings, an expansion of the $S$-matrix as
power series in matter fields has been used to obtain the moduli-dependence
of the coefficient of $\Phi_A^2$.  In writing (\ref{functs}) we made an
additional assumption of Heisenberg invariance, and dropped higher order terms
in the twisted sector fields. The latter cannot affect masses if
twisted sector fields vanish during inflation.  Under these assumptions,
the masses of the twisted sector fields during inflation can be determined: 
$m_A^2= F(q_I^A,b_A)V$, where $|F(q_I^A,b_A)|\sim 1$.  For example,
if we take $W_A = W_I = W_{A1} = W_{A2} = 0$ and assume $|W_{A3}| >> |W|$ 
(as in the explicit model introduced below), we get twisted sector masses 
$m_B^2 = V[1 - (1 + p_B\ell)q^B_3]$.
If Heisenberg invariance is not preserved by $K$ and $V_{GS}$, 
there can be additional terms of the form $X_A|\phi_{B_I}|^2/(t_I + \bar{t}_I)
$; these can modify the masses of the twisted fields by coefficients of order
one.  

{}From the above expressions we find
\be K_\alpha = \(\prod_I x_I^{-q^\alpha_I} \)\bar\phi_\alpha \,,\label{kalpha}
\ee
and near the origin
\be K_{\alpha\beta} = \(\prod_I x_I^{-q^\alpha_I} \)\delta_{\alpha\beta} \,.
\label{norigin}\ee

Supersymmetry is supposed to be broken by gaugino 
condensation. The condensates have masses larger than the condensation scale
$\Lambda_c = |u|^{1\over3}\sim 10^{13}\GeV$,
where $u$ is the $vev$ of the gaugino condensate. Below this scale,
we can integrate them out to get the effective theory.

The potential including matter fields was not given explicitly in \cite{bgw}. 
Assuming that the $D$ term vanishes, it is
\bea V &=& {1\over16\ell^2}\(\ell g'(\ell) + 1\)\left|u(1+b_a\ell) 
- 4\ell W(\phi)e^{K/2}\right|^2  \nonumber\\
& & - {3\over16}\left|b_au - 4W(\phi)e^{K/2}\right|^2  
+ \gh^{n\bar m} Y_n \bar Y_{\bar m} \,, \eea
where $b_a$ is one third the $\beta$-function coefficient for the confined
hidden gauge group ({\it e.g.}\/, $b_a =b$ for $E_8,\; b_a =
n/8\pi^2$  for pure $SU(n)$ Yang-Mills, {\it etc.}\/). 
The subscript $n$ takes on the values $I$ (corresponding to $t_I$), 
and $\alpha=A$ or $AI$. Roughly speaking
the first and last terms correspond, respectively, 
to the $F$-terms for the dilaton and for the other fields
in the usual chiral formulation, while the middle
term corresponds to the usual $-3e^K|W|^2$.

The factor $\gh^{n\bar m}$ is the inverse of the matrix $\gh_{n\bar m}$, 
with
\bea \gh_{n\bar m}  &=& G_{n\bar m} + \ell V\su{GS}_{n\bar m} \\
&=& (1+b\ell) \tG_{n\bar m} + \sum_A\left[ 1+ p_A\ell\right] (X_A)_{n\bar m}
\eea
and
$\xi=\eta'/\eta$ is the logarithmic derivative of the Dedekind function
$\eta$. The factors $Y_n$ are given by
\bea Y_n &=& e^{K/2} \left(W_n + G_n W\right)
+ \frac{u}{4}(b-b_a) \tG_n \nonumber \\
&&+\frac{u}{4}\sum_A(p_A-b_a) (X_A)_n - \frac{u}{2}(b-b_a)\xi(t_I)
\delta_{n I} \,.
\eea

Because we are dealing with a linear multiplet, the
superpotential $W$ is independent of the dilaton. This is in contrast with
the case for the chiral multiplet formulation, and is 
an  important simplification. The superpotential 
has a power series expansion in the 
matter fields which we take to be 
\be W=\sum_m \lambda_m \prod_\alpha \phi_\alpha^{n_m^\alpha} 
\prod_I\eta(t_I)^{2(\sum_\alpha n_m^\alpha q_I^\alpha -1)} \label{15} \ee
with the $n_m^\alpha$ positive integers or zero.
The $t_I$ dependence of each coefficient
is dictated by modular invariance, which requires that $W$
transforms like $\prod_I \eta^{-2(t_I)}$ (up to a
modular-invariant holomorphic function, which we do not consider
because it would have singularities). Using this expression one sees that
\be \frac{\pa W}{\pa t_I}\equiv
W_I  = 2\xi(t_I)\left( \sum_\alpha q^\alpha_I\phi_\alpha W_\alpha 
-W\right) . \ee

Putting all this together, the potential is
\bea V &=& {1\over16\ell^2}\(\ell g'(\ell) + 1\)\left|u(1+b_a\ell) - 4\ell
W e^{K/2}\right|^2 - {3\over16}\left|b_au -
4W e^{K/2}\right|^2   \nonumber\\
&&+ \sum_A \(\prod_I x_I^{q_I^A} \) \frac{|Y_A|^2}{1+p_A\ell}+
\sum_I{1\over1 + b\ell + \sum_B(1+p_B\ell)q^B_I X_B}\times \nonumber
\\ & & \[\left| A_I\left(2\xi(t_I)x_{I} +1\right)
- e^{K/2}\sum_A\phi_{AI}W_{AI}\right|^2 \right. \nonumber\\
&& \left. + x_{I}\sum_A\left|W_{AI}e^{K/2} + 2\xi(t_I)A_I\bp_{AI}\right|^2\].
\label{17} \eea
In this expression,
\be A_I = e^{K/2}\(\sum_\alpha q^\alpha_I\phi_\alpha W_\alpha - W \) - 
{u\over4}(b-b_a) , \ee
or equivalently
\be 2\xi(t_I) A_I = e^{K/2} W_I -2\xi(t_I){u\over4}(b-b_a) .\ee
Also,
\bea Y_A &=& {1\over\phi_A}\lbr e^{K/2}\[\phi_AW_A +
X_AW  \] + {u\over4}(p_A -b_a)X_A\rbr  \\
&=&  e^{K/2}\[W_A + K_A W \] + {u\over4}(p_A -b_a)K_A \eea
where $K_A = \(\prod_I x_I^{-q_I^A} \)
\bar \phi_A$.

This potential has degenerate vacua with broken supersymmetry,
at $t_I=1$ and $t_I= e^{i\pi/6}$ (with all matter fields equal to zero
in both cases).\footnote
{These vacua correspond to $2\xi(t_I)x_{I} +1=0$. Points
in field space that are obtained from these vacua by a 
modular transformation do not represent physically distinct vacua,
since the group of modular transformations is a gauge
discrete symmetry as opposed to a global one.}
Only the
former was considered in~\cite{bgw}, but the qualitative features are the same 
if one takes $t_I= e^{i\pi/6}$.  

If $V\gg u^2 $ (restoring the Planck mass, $V^{1/4}\gg 10^{11}\GeV$),
then $u$ is presumably negligible and we obtain 
\bea V &=& e^K\Bigg\{\(\ell g'(\ell) + 1\)|W|^2 -3|W|^2 + 
\sum_A\(\prod_I x_I^{q_I^A}\)\frac{\left| W_A + K_A W\right|^2}{1+p_A\ell}
\nonumber \\ && +
\sum_I{1\over1 + b\ell + \sum_B(1+p_B\ell)q^B_I X_B}\times\nonumber \\ 
& & \[\left|x_I W_I - W +\sum_A q^A_I \phi_A W_A \right|^2  
+ x_{I}\sum_A\left|W_{AI} + W_I \bp_{AI}\right|^2\]\Bigg\}\,. \label{17x} \eea
This corresponds to \eq{9}, with the dilaton and twisted-sector fields
now included.

\section{Building a model of inflation}

Guided by references \cite{cllsw,ewansg}, we suppose that during 
inflation the following conditions hold.
\begin{enumerate}
\item The gaugino condensate $u$ is negligible, corresponding to
$V^{1/4}\gg \sqrt u \sim 10^{11}\GeV$.
\item Every term in the expansion (\ref{15}) of $W$ vanishes
(i.e., at least one of the fields in each term vanishes).
\item All derivatives of $W$ with respect to the fields vanish,
except for $W_{C3}$ which is fixed during inflation corresponding to a single
untwisted matter field $\alpha=C3$.\footnote
{The choice $I=3$ is arbitrary, and one 
could allow nonvanishing $W_\alpha$ for more untwisted 
fields from the {\em same} sector without changing anything.
As we shall see later, one might also allow nonvanishing $W_\alpha$ 
for fields from two or all three of the untwisted sectors, but for the 
moment we insist on just one.}
\item All matter field values are $\ll 1$.
\end{enumerate}
Somewhat less specific conditions would have essentially the same effect, but
these have the virtue of simplicity, and we shall later 
be making a specific
proposal for achieving them. 

Since the twisted fields $\phi_A$
are $\ll 1$, the terms $X_A$ defined by \eq{xa} are also $\ll 1$.
Both $\phi_A$ and $X_A$ can
be ignored in \eq{17x}, which 
becomes simply
\be V = \frac{\ell e^{g(\ell)}}{(1+b\ell) x_1 x_2}  
|W_{C3}|^2 \,.
\label{26} \ee

\subsection{A simple possibility}

In order to proceed, we need to know the dependence
of $|W_{C3}|$ on the moduli and the dilaton.
Following reference \cite{cllsw}, let us first
suppose that $W_{C3}$ (considered as a function of the matter fields 
and the $t_I$) is independent of $t_3$, but has dependence on $t_1$
and $t_2$ which stabilizes the potential. Then
flat directions 
are preserved for matter fields in 
the $I=3$ sector 
(except for any which are spoiled by coupling to fields that are
displaced from the origin)
and the $t_3$ direction is also flat.
This is because the terms in \eq{17x} that give zero contribution to 
$V$, also give zero contribution to the squared masses of these fields.
Flat directions in the $I=1$ and $I=2$ sectors are not preserved
because of the factors $x_1$ and $x_2$ in \eq{26}. 
As noted earlier, the masses of twisted sector fields obtained from
\eq{17x} could be modified by unknown coefficients of order one, if the
assumption of Heisenberg invariance of the K\"ahler potential is dropped.
The inflaton trajectory could be any combination of $t_3$ and the
flat $I=3$ directions
(excluding $\phi_{C3}$ which is supposed to be fixed).

These conditions would be achieved \cite{cllsw} if
$W_{C3}$ came from a term $\Lambda^2 \phi_{C3}$,
with $\Lambda$ independent of the matter fields.
Then, modular invariance would imply that
$\Lambda^2\propto \eta^{-2}(t_1) \eta^{-2}( t_2)$, and 
\be V\propto \left[|\eta (t_1) \eta( t_2)|^4 x_1 x_2
\right]^{-1} \,.
\label{vmod} \ee
To discuss the stability of the moduli, we can set the matter fields 
equal to zero so that  $x_I=t_I+\bar t_I$.
As shown in \cite{cllsw}, $V$ is stabilized at
$t_1=t_2=e^{i\pi/6}$ up to modular 
transformations. The squared masses of the canonically normalized
$t_1$ and $t_2$ turn out to be precisely $V$, which presumably hold them in 
place during inflation. 

The value $t_I= e^{i\pi/6}$ corresponds to a fixed point\footnote{The other
fixed point in the fundamental domain, namely $t_I = 1$, is a saddle point of 
potential (\ref{vmod}); see e.g.~\cite{font}.}
of the modular transformations. Since it must be an extremum of the 
potential, it is not particularly surprising to find that it represents
the minimum during inflation.  As we noted earlier it also represents a 
possible true vacuum. As a result, the moduli stabilized at this point
during inflation will remain there, and, as has often been noted 
before,  would not be produced in the early Universe.

To complete this simple model, note that $|W_{C3}|$ has no
dependence on the dilaton. The \eq{26} gives
\bea V &=& \lambda {\ell e^g\over1+b\ell}, \quad V' = {V\over\ell}\(1 + \ell g' 
- {b\ell\over1+b\ell}\), \nonumber \\ 
V'' &=& {V\over\ell^2}\(\ell^2g'' - 1 + {b^2\ell^2\over(1+b\ell)^2}\) + 
{V'\over\ell}\(1 + \ell g' - {b\ell\over1+b\ell}\). \eea
We require $V' = 0, \;V'' > 0$ for stabilization, which means 
\be \ell g' + 1 = {b\ell\over1+b\ell},  \quad \ell^2g'' > 1 - 
{b^2\ell^2\over(1+b\ell)^2}.\ee
The function $f(\ell)$ is related to $g(\ell)$ 
by a differential equation (which assures a canonical form for the
Einstein term in the Lagrangian):
\be g'\ell = f - f'\ell. \ee 
An example (taken here for calculational simplicity) of a choice for $f$ 
that reflects string nonperturbative effects~\cite{shenk}, and stabilizes
the dilaton at weak coupling [$\alpha(m_{str})\approx$ .17] and vanishing 
cosmological constant, is 
\be f = 4.25e^{-1/\sqrt{b\ell}}(1 - .53/\sqrt{b\ell})\,.\label{fexp}\ee
With this parameterization the potential (\ref{26}) has a local 
minimum at $\ell = 4.2$, which is 
in the domain of attraction and roughly satisfies our initial assumption that 
$\ell= O(1)$ during inflation. 

\subsection{More general possibilities}

Contrary to what was stated in \cite{cllsw}, one cannot
argue that \eq{vmod} always holds, because in general $W$ is an arbitrary
expression of the form \eq{15}. Although $W_{A3}$ 
transforms like $(\eta(t_1)\eta(t_2))^{-2}$, this is automatically
satisfied by \eq{15} and it does not in general 
determine  the dependence of $W_{A3}$ on the $t_I$ at (say) fixed values
of the matter fields.  More specifically, 
since we are working in the context of
string theory, we need to justify the emergence of a superpotential 
term linear 
in a matter field, since the effective Lagrangian from string theory contains
terms of cubic and higher order.  Thus a linear term
can arise only from some fields
acquiring $vev$'s.

In fact, something like \eq{vmod} may be applicable
under rather general circumstances. Let us a assume that
at the string
scale, $W$ includes a term of the form 
\be W = \lambda \phi_{C3}\[\eta(t_1)\eta(t_2)\]^{-2}
\prod_{\alpha}\phi_\alpha
\prod_I\[\eta(t_I)\]^{2q_I^\alpha}\,, \label{W} \ee
 where the product over $\alpha$ does not contain $\phi_{C3}$.
We suppose that during inflation, there are nonzero vevs 
$|\phi_\alpha|^2$, with the 
modular invariant form
\be |\phi_\alpha|^2\prod_I x_I^{-q^\alpha_I} = c_\alpha
\ell^{d_\alpha}\prod_I\[x_I|\eta(t_I)^4|\]^{p^\alpha_I},
\label{vevs}\ee
where $c_\alpha$ is a constant. 
As we shall see, $vev$'s of this form can indeed be generated from 
a Fayet-Illiopoulos
$D$ term. Such vevs will drive $\phi_{C3}$ to zero as required.
Once the fields with these $vev$'s
have been integrated out to give an effective theory relevant to the scale of
inflation, the moduli-dependence of the potential~(\ref{26}) takes the form
\be
V\propto \prod_I\left[ |\eta(t_I)|^4 x_I \right]^{n_I}, \quad
n_{I= 1,2} = \sum_\alpha\(p_I^\alpha + q_I^\alpha\) - 1,\quad
n_3  = \sum_\alpha\(p_3^\alpha + q_3^\alpha \)\,.
\label{modpot}\ee

Bearing in mind our earlier discussion, we
want one or more of the
$n_I$ to vanish, providing flat directions suitable for inflation.
Any remaining $n_I$ should be negative, which as we noted
after \eq{vmod} will ensure that
the corresponding moduli
are stabilized at $t_I = e^{i\pi/6}$.
(Positive $n_I$ are excluded, because the potential would be 
driven to zero in the direction $t_I\to\infty$ (or 0).)
Inflaton candidates are the modulus (or moduli) with 
$n_I=0$, and matter fields in the corresponding untwisted sector(s)
which correspond to flat directions. 

In contrast with the earlier situation, any or all of
the $n_I$ can vanish. If they all vanish, one could generalize \eq{W} to 
be a sum over terms, with $\phi_{C3}$ is replaced by fields from
{\em different} untwisted sector. 

The dilaton-dependence of $V$ is
\be V\propto {e^{g(\ell)}\ell^d\over 1 + b\ell},\quad d = 1 +  \sum_\alpha 
d_\alpha , \label{dilpot} \ee
for which the minimization conditions (32) become
\be \ell g' + d = {b\ell\over1+b\ell},  \quad \ell^2g'' > d - 
{b^2\ell^2\over(1+b\ell)^2}.\label{min}\ee
The condition that the potential be positive definite requires~\cite{bgw,bgw1}
$\ell g' > - 1$ and since $0\le b\ell/(1+b\ell) <1$,  stabilization can occur
only if $d < 2$.  With the parameterization introduced above, there is a
minimum within the domain of attraction ({\it i.e.}, with $\ell\gtrsim
1.4$) for $-3.3\lesssim d\lesssim 1.4$.

Taken literally, this model gives an exactly flat inflaton potential,
and no mechanism for ending inflation. There are many possibilities
for generating a slope. It can come 
from small departures from assumptions 2--4,
from the gaugino condensate or loop corrections. 
Also, if the $vev$'s are generated by a $D$ term
that term will be driven to a small but nonzero value; this
will generate a slope from the inflaton-dependence of the
factors $K_{n\bar n}$ in \eq{dterm98}.  With the slope in place, 
a generalization of the model exhibited in reference \cite{ewansg} 
allows inflation to end by the hybrid inflation mechanism.

Provided that no
matter fields charged under the strongly coupled hidden gauge group
acquire large $vev$'s during inflation, the condensate potential $V_c$
will indeed be present. Let us estimate the mass it generates for 
the moduli with $n_I=0$. 
As mentioned above, the model is viable only if the moduli are close to their
vacuum values -- {\it i.e.} within the domain of attraction -- during inflation.
In this case the mass of moduli with $n_I=0$ is 
$m^2_{t_I} \approx 
V_c(\ell_i)$, where $\ell_i$ is the value of the dilaton field during
inflation.  The magnitude of $V_c$ is governed by the value of the 
string-scale gauge coupling $g^2(\ell_i) = 2\ell_i/[1+f(\ell_i)]$,
and the condition that the vacuum energy vanishes in the true vacuum assures
that this is a slowly varying function near its vacuum value $\ell_v$.  For the 
parameterization used above with $d=1$, $\(g^{-2}(\ell_v),g^{-2}(\ell_i)\) = 
(.44,.13),\; u(\ell_i) \approx 10^2u(\ell_v),\; m_{t_I}(\ell_i) 
\approx 5m_t(\ell_v)\approx 100\TeV \ll V^{1\over2}$.
(Note that $V_c(\ell_i)$ is of order $(10^{11}\GeV)^4$ 
as one would expect.)

Flat directions in the corresponding untwisted sector are lifted by 
mass terms of order $|m_{\phi_{AI}}|^2 \sim m_{t_I}^2$ as long as 
$|\phi_{IA}|^2\lesssim {\re t_I}$. This contribution is negative if 
$|\phi_{IA}|^2\lesssim .2
{\re t_I}$, and is much smaller than that induced~\cite{gmo} by loop effects 
($-m^2_\phi \sim 10^{-2}V$). If either of these gives the dominant slope, the
spectral index $n=1+2m^2/V$ is very close to 1.

Finally, we note that the ``moduli problem''~\cite{modprob,lsp} encountered in 
generic supergravity/superstring inflationary scenarios, is considerably
alleviated in the model studied here.  While the dilaton is stabilized 
at a value shifted from its vacuum value by an amount of $O(1)$, 
its mass is about $10^6GeV$~\cite{bgw}, and its decay does not
contribute to the moduli problem.\footnote{Even though there is no
  constraint from the Big-Bang Nucleosynthesis because of its high
  mass and hence early decay, it still dilutes the baryon asymmetry by
  a factor of roughly $10^{-12}$.
  A very efficient mechanism, such as Affleck--Dine mechanism, can
  generate enough baryon asymmetry to withstand the dilution \cite{MYY}.} 
The moduli masses are about 20$\TeV$, which is sufficient to
evade the moduli problem of~\cite{modprob} only if R-parity is 
violated~\cite{lsp} (or the moduli abundance is diluted by thermal
inflation \cite{thermal}).
If R-parity is conserved, the problem is still evaded for those moduli that 
are stabilized at the vacuum value $t_I=e^{i\pi/6}$.  It is possible that the
requirement that the remaing moduli ({e.g., $t_3$ in the above example)
be in the domain of attraction is sufficient to avoid the problem
altogether. 

\subsection{Generating $vev$'s with the $D$ term}

In many models $vev$'s of the form (\ref{vevs}) with $p^\alpha_I=0,\; d_\alpha 
= 1$ arise from a Fayet-Illiopoulos 
$D$ term, whose contribution to the potential is given by \eq{dterm98}.
In string models
it arises as a GS counter term, introduced\cite{dsw} to cancel a $U(1)$ gauge 
anomaly of the effective field theory (with no corresponding string theory 
anomaly), analogous to the the GS term introduced in III.B to cancel the modular
anomaly. Orbifold models with an anomalous $U(1)$ and supersymmetric vacua
have been found in~\cite{iban,font2}, and the GS 
$D$ term has been used in various applications to phenomenology.  

In the linear multiplet formalism, the
gauge coupling constant $g$ (defined at the string scale)
which appears in \eq{dterm98} is given by
\be g^2 = \frac{2\ell}{f(\ell) + 1} \,. \label{g2} \ee 
The scale $\xi_D$ of the Fayet-Illiopoulos term is given in this formalism
by
\be\xi_D = \frac{2\ell \Tr(T)}{192\pi^2} \,,\label{xiofg}\ee
where $T$ is the generator of the anomalous $U(1)$, whose trace $\Tr(T)=\sum
q_n$ is perhaps \cite{kobayashi96,font2} of order 100. 

Using \eq{kalpha}, one sees that $vev$'s generated by the $D$ term
will be of the form \eq{vevs}.\footnote
{ In \cite{ewansg}, $K_{n\bar m}$ was set equal to $\delta_{nm}$
in the $D$-term.
Including the nontrivial $K$ will in general affect the flatness,
as the present discussion demonstrates.}
If this were the only source of
$vev$'s, $d$ would just be the dimension of the the superpotential in (\ref{W})
with $d \ge 3$. The potential (\ref{dilpot}) would be driven to zero in the
direction of vanishing gauge coupling $\ell\to 0$.  However $vev$'s induced by 
a $D$ term can induce other $vev$'s with a different $\ell$-dependence through 
superpotential terms.  If, for example, there is a gauge invariant, modular 
covariant superpotential term 
\be w = F\prod_{\beta=1}^3\phi_\beta\prod_I[\eta(t_I)]^{2q_I^\beta},
\quad F = \prod_I\eta^{-2}(t_I),\label{w2} \ee
the superpotential
\be W(w) = F\sum_n c_n(w/F)^n, \ee
is allowed by all the symmetries.  Now suppose the $D$ term induces
modular invariant $vev$'s for $\phi_2$ and $\phi_3$:%
\footnote{The ratio of $\phi_2$ and $\phi_3$ $vev$'s can be
fixed, for instance, by gauging a non-anomalous U(1) symmetry under
which they have the opposite and equal charges.}
\be |<\phi_\beta>|^2 = |v_\beta|^2 = c_\beta\ell
\prod_I(t_I + \bar{t_I})^{q_I^\beta}, \quad \beta = 2,3.\ee
Then if $c_1$ is nonzero, $W_1(w)$ does not vanish unless $<\phi_1> = v_1 \ne 
0.$ Solving $W_\beta = 0, \;\beta = 1,2,3$ gives a single
equation: $\phi^\beta W_\beta = \sum_n n c_n(w/F)^n = 0$, which  is solved by
$w/F = \sqrt{c_1} =$ constant.  Then 
\be |v_1|^2 = {c_1\over c_2c_3}\ell^{-2}\prod_I x_I^{q^1_I} 
\[x_I|\eta(t_I)|^4\]^{p^1_I}, \quad p^1_I =
\sum_{\beta=1}^3q^\beta_I.\ee
It is easy to satisfy the condition $d<2$ by including such fields in the
superpotential (\ref{W}).  For example, if $w = \phi_{A1}\phi_{B2}\phi_{C'3}$
in (\ref{w2}) and $W = \phi_{C3}\phi_{A1}\phi_{B'2}\phi_{C''3}\eta(t_3)^2$ 
in (\ref{W}), with $D$ term induced $vev$'s for $\phi_{B2},\phi_{C'3},
\phi_{B'2},\phi_{C''3}$, one recovers precisely the behavior in (\ref{26})
and (\ref{vmod}).

The magnitude of the potential will be of the form
\be V= \lambda \Lambda^{-2n}\xi_D^{2(2+n)} .\ee
In this expression, $\lambda$ is a ratio of dimensionless couplings in the 
superpotential (times a coefficient of order 1), 
$3+n$ is the dimension of the term \eq{W} of the 
superpotential, and $\Lambda$ is
scale of nonrenormalizable terms in the superpotential.
Using \eq{xiofg}, and the perturbative superstring estimate
$\Lambda^2 = M\sub{str}^2\equiv g\sub{str}^2 \mpl^2$,
this gives 
\be
V^{1/4}\sim 
\lambda^{1/4}g \sub{str}\left(\frac{\Tr(T)}{192\pi^2}\right)^{(2+n)/4} 
\mpl 
\,.
\label{vofmstring}
\ee

One expects that 
$g\sub{str}$
will be at most an order of magnitude 
below unity,
\footnote
{$g^2\simeq 0.5$ would correspond 
to the value $\alpha\sub{str}\equiv g\sub{str}^2/4\pi
\simeq 1/25$, which with naive running of the couplings is suggested
by observation at a scale of order $10^{16}\GeV$. 
In fits (including additional particles) to the string theory constraint
that unification occurs at $\mu^2= (2e)^{-1} g^2\mpl^2$, 
the value of $g^2$ is somewhat larger.} 
but the 
other factors
can be smaller. With reasonable
values like $n=1$ or $n=2$, one can easily achieve the result
$V^{1/4}\lsim
10^{-2}\mpl$ required by the COBE normalization.

\section{Conclusion}

We have shown how to construct a general class of inflation
models, with some very desirable properties.
The scale of inflation is set by a Fayet-Illiopoulos term, derived
from the superstring.
Some of the flat directions of global SUSY are preserved, 
and the potential is also 
flat in the direction of at least one of the moduli. All of these flat 
directions are candidates for the inflaton field.
The dilaton is stabilized within its domain 
of attraction, and the remaining moduli are stabilized
at or near their vacuum values. The models may avoid the usual 
moduli problem.

These inflation models are constructed within the framework of a specific, 
modular invariant, model of supersymmetry breaking in the true vacuum, 
that invokes string nonperturbative effects to stabilize the dilaton. 
It is based on a class of orbifold compactifications with just three untwisted 
moduli $t_I$, and the dilaton is described 
by the linear supermultiplet formalism.  

Although the specific model contains a definite mechanism for stabilizing the
dilaton in the true vacuum, this is actually irrelevant for our
proposed models of
inflation because they make 
the inflationary energy scale much bigger than the 
scale of SUSY breaking in the true vacuum ($V^{1/4}\gg 10^{11}\GeV$).
If, for instance, SUSY breaking in the true vacuum is gauge-mediated,
our models of inflation still work provided that the dilaton
is described by the assumed linear-multiplet formalism.

Let us emphasize that what we have given is only a 
strategy for model-building. We have shown how to achieve a sufficiently 
flat inflationary potential, but we have not exhibited a complete model.
Such a model would define the slope of the inflationary potential,
and would include a mechanism for ending inflation. To achieve this last 
objective we would presumably need a hybrid inflation model, along the 
lines of the one given in reference \cite{ewansg}. To exhibit such a 
model would seem to be a worthwhile project, and might be
quite nontrivial.

\subsection*{Acknowledgements}

MKG and HM were supported in part by the Director, Office of Energy
Research, Office of High Energy and Nuclear Physics, Division of High
Energy Physics of the U.S. Department of Energy under Contract
DE-AC03-76SF00098 and in part by the National Science Foundation under
grant PHY-90-21139.  HM was also supported by the Alfred P. Sloan
Foundation.  DHL is supported by PPARC grant GR/L40649 and NATO 
grant CRG 970214, and acknowledges hospitality at LBNL where this work 
was started. DHL acknowledges a long-term collaboration with Ewan
Stewart, and useful correspondence with Chris Kolda and Toni Riotto.
MKG thanks Yi-Yen Wu for discussions.

\end{document}